\documentstyle[11pt]{article}
\begin{document}
\renewcommand{\theequation}{\thesection.\arabic{equation}}
\renewcommand{\refname}{References.}
\newcommand{\sect}[1]{ \section{#1} \setcounter{equation}{0} }
\newcommand{\partialslash}{\partial \! \! \! /}
\newcommand{\pslash}{p \! \! \! /}
\newcommand{\kslash}{k \! \! \! /}
\newcommand{\lslash}{l \! \! \! /}
\newcommand{\qslash}{q \! \! \! /}
\newcommand{\Dslash}{D \! \! \! /}
\newcommand{\xslash}{x \! \! \! /}
\newcommand{\yslash}{y \! \! \! /}
\newcommand{\zslash}{z \! \! \! /}
\newcommand{\cosech}{\mbox{cosech}}
\newcommand{\half}{\mbox{\small{$\frac{1}{2}$}}}
\newcommand{\halfsmall}{{\mbox{\footnotesize{$\frac{1}{2}$}}}}
\newcommand{\third}{\mbox{\small{$\frac{1}{3}$}}}
\newcommand{\quarter}{\mbox{\small{$\frac{1}{4}$}}}
\newcommand{\sixth}{\mbox{\small{$\frac{1}{6}$}}}
\newcommand{\eighth}{\mbox{\small{$\frac{1}{8}$}}}
\newcommand{\twothirds}{\mbox{\small{$\frac{2}{3}$}}}
\newcommand{\threehalves}{\mbox{\small{$\frac{3}{2}$}}}
\newcommand{\fourthirds}{\mbox{\small{$\frac{4}{3}$}}}
\newcommand{\sixteenth}{\mbox{\small{$\frac{1}{16}$}}}
\newcommand{\la}{\langle}
\newcommand{\ra}{\rangle}
\newcommand{\Nf}{N_{\!f}}
\newcommand{\N}{\mbox{\small{$\frac{1}{N}$}}}
\newcommand{\R}{\frac{C_2(R)}{T(R)}}
\newcommand{\M}{\frac{1}{M}}
\newcommand{\G}{\frac{C_2(G)}{T(R)}}
\newcommand{\Pio}{{\Pi^{\mbox{o}}\!\!\!}}
\title{On the $O(1/N^2)$ $\beta$-function of the Nambu--Jona-Lasinio
model with non-abelian chiral symmetry.}
\author{J.A. Gracey, \\ Department of Applied Mathematics and Theoretical
Physics, \\ University of Liverpool, \\ P.O. Box 147, \\ Liverpool, \\
L69 3BX, \\ United Kingdom.}
\date{}
\maketitle
\vspace{5cm}
\noindent
{\bf Abstract.} We present the formalism for computing the critical exponent
corresponding to the $\beta$-function of the Nambu--Jona-Lasinio model with
$SU(M)$ $\times$ $SU(M)$ continuous chiral symmetry at $O(1/N^2)$ in a large
$N$ expansion, where $N$ is the number of fermions. We find that the equations
can only be solved for the case $M$ $=$ $2$ and subsequently an analytic
expression is then derived. This contrasting behaviour between the $M$ $=$ $2$
and $M$ $>$ $2$ cases, which appears first at $O(1/N^2)$, is related to the
fact that the anomalous dimensions of the bosonic fields are only equivalent
for $M$ $=$ $2$.

\vspace{-18cm}
\hspace{10cm}
{\bf {LTH-314}}
\newpage
\sect{Introduction.}
Quantum field theories with a four fermi interaction are known to possess an
interesting range of properties. For example, the two dimensional model,
introduced by Gross and Neveu in \cite{1}, is an asymptotically free theory
where the fundamental particles of the massless classical theory acquire a mass
dynamically in the quantum theory. Therefore, it mimics several of the features
of more involved four dimensional field theories such as quantum chromodynamics
and so has been used as a simple laboratory for testing out ideas which are
harder to examine there. More recently, four dimensional models possessing a
four fermi interaction have received intense study following the observation of
Nambu, \cite{2}, that such models could provide a realistic alternative to the
Higgs boson for generating a mass for the particles seen in nature, \cite{3,4}.

One technique which can be used to analyse models with a four fermi interaction
is the large $N$ expansion, \cite{1}, where $N$ is the number of fermions. For
instance, computing the effective potential of the two dimensional model in the
saddle point approximation when $N$ is large one observes that the perturbative
vacuum of the theory is not the correct one but that where the vacuum
expectation value of a bosonic auxiliary field is non-zero which in turn
generates the fermion mass. This behaviour is also preserved in the three
dimensional model and it is important to note this property is non-perturbative
and therefore would never be accessed in a perturbative approach. However,
whilst the large $N$ expansion has been powerful in revealing such structure
the conventional method possesses one particular drawback in that it is
virtually impossible to calculate beyond the leading order. This is primarily
due to the nature of the (dynamically generated) propagators of the bosonic
$\bar{\psi}\psi$ bound states which exist in the model and are non-fundamental
in structure, \cite{1}. Thus computing Feynman integrals with such massive
propagators becomes impossible beyond leading order.

Clearly there is a need to probe such models to higher orders in $1/N$ and to
have a technique which interpolates between the spacetime dimensions already
discussed. One such method currently does exist which achieves these aims is
the critical point self consistency method of Vasil'ev et al which was
developed for the $O(N)$ bosonic $\sigma$ model in \cite{5,6}. The beauty of
that approach is that one calculates within the theory at the fixed point which
is defined as the non-trivial zero of the $d$-dimensional $\beta$-function. At
this point the theory is conformal and therefore massless which allows one to
overcome the difficulty of computing previously intractable integrals beyond
the leading order. Moreover, one solves the model by determining the critical
exponents of the fields and Green's functions \cite{5}. These can then be
related to the perturbative renormalization group functions through an
examination of the renormalization group equation in the critical region.
Examples of this approach can be found in \cite{7}. Further, the
$\beta$-function of any theory carries all the important information on the
evolution of the coupling constant with the renormalization scale, and it too
can be computed order by order in $1/N$ by considering the appropriate critical
exponent, \cite{6}. As these are determined as functions of the spacetime
dimension one can deduce information on models simultaneously in several
different dimensions.

In this paper, we complete the examination of the four fermi models at
$O(1/N^2)$ by computing the $\beta$-function exponent of the
Nambu--Jona-Lasinio model with a non-abelian continuous global chiral symmetry.
Previously various exponents had been calculated in the $O(N)$ Gross Neveu
model in \cite{8,9,10,11} using the original large $N$ approach. The
breakthrough to $O(1/N^2)$ was achieved in \cite{12} with the computation of
the fermion anomalous dimension $\eta$ and latterly the mass, \cite{13,14},
vertex, \cite{13,14}, and $\beta$-function exponents, \cite{14,15} have been
determined as well as $\eta$ at $O(1/N^3)$, \cite{14,16}. Several of these
exponents were subsequently determined to $O(1/N^2)$, \cite{17}, for the
generalization of the $O(N)$ model, which possesses a discrete chiral symmetry,
to the case where it has a $U(1)$ $\times$ $U(1)$ or $SU(M)$ $\times$ $SU(M)$
continuous global chiral symmetry, \cite{1,18}. Several leading order results
had earlier been presented in \cite{11,19} using the saddle point approach. By
computing the exponent $\lambda$, where $2\lambda$ $=$ $-$ $\beta^\prime(g_c)$,
at $O(1/N^2)$ in the $SU(M)$ $\times$ $SU(M)$ case here the model can then be
said to have been solved thermodynamically since knowledge of two independent
exponents means that the remaining ones can be deduced through hyperscaling
laws which have been checked at leading order in \cite{11} and their
consistency merely reflects the renormalizability of the model. Several
additional motivations for considering the $SU(M)$ $\times$ $SU(M)$ model in
particular include the provision of analytic results for the three dimensional
model which will provide key estimates for numerical simulations of the model
on the lattice. Recently, estimates for several exponents have been determined
using Monte Carlo methods in the $O(N)$ Gross Neveu model for relatively small
values of $N$ in \cite{20} and they are in agreement with $O(1/N^2)$ results,
\cite{15}. Further the provision of expressions valid in $d$-dimensions as a
function of $N$ is important for demonstrating the equivalence of various
models. For example, it is known that the $(4-\epsilon)$-dimensional Yukawa
model and the $(2+\epsilon)$-dimensional Gross Neveu model lie in the same
universality classes which has been established by examining the
$\epsilon$-expansions of the perturbative renormalization group functions to as
many orders as they are presently known. In our case the equivalence of the
$(2+\epsilon)$-dimensional non-abelian Nambu--Jona-Lasinio or chiral Gross
Neveu model will probably be with the generalized $(4-\epsilon)$-dimensional
Gell--Mann-L\'{e}vy $\sigma$ model introduced in \cite{21} and currently of
interest due to its relation to hadronic physics. (The original model possessed
an $SU(2)$ $\times$ $SU(2)$ chiral symmetry and by generalized we mean its
natural extension to the case of $SU(M)$ $\times$ $SU(M)$.) The provision of
critical exponents to $O(1/N^2)$ which we carry out here will be the first step
in such a proof.

Finally, we mention that there is some uncertainty in the literature concerning
the behaviour of the $SU(M)$ $\times$ $SU(M)$ model for $M$ $>$ $2$ and for $M$
$=$ $2$, \cite{11}. It has been suggested in \cite{11} that the chiral symmetry
is only realised in the case $M$ $=$ $2$ and not for higher $M$, \cite{1}. So
far in the large $N$ self consistency approach there has been no indication of
distinct behaviour between either case. This is normally observed in the
breakdown of the computation of the $\beta$-function exponent at some order in
large $N$ which has been discussed in the context of other models in \cite{22}.
At leading order we were able in \cite{17} to compute $\eta$, $\lambda$ and the
vertex anomalous dimensions, $\chi$, at $O(1/N)$ and $\eta$ and $\chi$ at
$O(1/N^2)$ without complications for all $M$. So it will be interesting to see
if it is possible to deduce $\lambda$ at $O(1/N^2)$ for all $M$.

The paper is organised as follows. In section 2 we introduce the details of
the model we will examine at criticality and illustrate the method by
computing $\eta$ at $O(1/N)$. The leading order analysis is continued in
section 3 where the calculation of $\lambda$ at $O(1/N)$ is detailed in
preparation for the $O(1/N^2)$ analysis which is presented in section 4. We
conclude the paper by solving the master equation in section 5 where we
discuss the results for $M$ $=$ $2$ and $M$ $>$ $2$ separately.

\sect{Preliminaries.}
To begin with we describe the model we are interested and introduce the
notation and formalism we will use in solving it. Its lagrangian can be written
in several ways but we choose to use the auxiliary field version since it
involves three point vertices which are essential to the technique of
uniqueness used to compute the Feynman graphs, \cite{23}. Thus we take
\cite{1,18}
\begin{equation}
L ~=~ i \bar{\psi}^{iI} \partialslash \psi^{iI} + \sigma \bar{\psi}^{iI}
\psi^{iI} + i \pi^a \bar{\psi}^{iI} \lambda^a_{IJ}\gamma^5 \psi^{iJ}
- \frac{1}{2g^2}(\sigma^2 + \pi^{a2})
\end{equation}
where $\psi^{iI}$ is the fermion field with $1$ $\leq$ $i$ $\leq$ $N$, $1$
$\leq$ $I$, $J$ $\leq$ $M$, $1$ $\leq$ $a$ $\leq$ $(M^2-1)$ and $g$ is the
coupling constant. Eliminating the auxiliary bosonic fields $\sigma$ and
$\pi^a$ from (2.1) would yield the four fermi interaction explicitly and
according to \cite{1} (2.1) possesses an $SU(M)$ $\times$ $SU(M)$ continuous
global chiral symmetry. The matrices $\lambda^a_{IJ}$ are the generalized Pauli
matrices of the group $SU(M)$, where $[\lambda^a,\lambda^b]$ $=$
$2if^{abc}\lambda^c$ and $f^{abc}$ are its structure constants, and we
normalize them in as general a way as possible by taking
\begin{equation}
\mbox{Tr} (\lambda^a \lambda^b) ~=~ 4T(R) \delta^{ab}
\end{equation}
The Casimirs which will appear throughout the paper are
\begin{equation}
\lambda^a \lambda^a ~=~ 4C_2(R) I ~~,~~ f^{acd} f^{bcd} ~=~ C_2(G)
\delta^{ab}
\end{equation}
where for $SU(M)$, $T(R)$ $=$ $\half$, $C_2(R)$ $=$ $(M^2-1)/2M$ and $C_2(G)$
$=$ $M$. Also, in order to carry out intermediate checks on our calculation we
will retain $C_2(R)$ and $C_2(G)$ in their general form to allow us to take the
limits to either the case with $U(1)$ $\times$ $U(1)$ chiral symmetry,
($C_2(R)$ $=$ $T(R)$ $=$ $\quarter$, $C_2(G)$ $=$ $0$, $M$ $=$ $1$) or the
$O(N)$ model with discrete chiral symmetry, ($C_2(R)$ $=$ $C_2(G)$ $=$ $0$) and
compare with results we already know.

We now introduce the basics to enable us to solve (2.1) to $O(1/N^2)$ in the
critical point self consistency approach. First, the fundamental idea is to
examine the scaling behaviour of the fields of (2.1) in the critical region
defined as the non-trivial zero of the $\beta$-function in $d$-dimensions,
\cite{5,6,12}. Since the fields are massless at this fixed point where there is
a conformal symmetry the structure of the Green's functions take a particularly
simple form. For instance, in coordinate space they are
\begin{equation}
\psi(x) ~ \sim ~ \frac{A\xslash}{(x^2)^\alpha} ~~,~~
\sigma(x) ~ \sim ~ \frac{B}{(x^2)^\beta} ~~,~~
\pi(x) ~ \sim ~ \frac{C}{(x^2)^\gamma}
\end{equation}
as $x^2$ $\rightarrow$ $0$ where we use the same letter to denote the
propagator and $A$, $B$ and $C$ are $x$-independent amplitudes. By analogy with
ideas in statistical mechanics the properties of the model in the critical
neighbourhood are described totally by the critical exponents $\alpha$, $\beta$
and $\gamma$ of each of the individual fields, \cite{7}. By the universality
principle they are functions only of the spacetime dimension, $d$ $=$ $2\mu$,
and any internal parameters of the theory. For our case, these will be $N$ and
$M$ through the appearance of the Casimirs of (2.3). Moreover, the exponents
can be related to the appropriate renormalization group functions which are
ordinarily calculated perturbatively as a series in $g$, by examining the
renormalization group equation at criticality, \cite{7}. For instance, if we
examine the canonical dimension of each field in the action which is a
dimensionless quantity then we can define their anomalous dimension as follows,
\cite{17},
\begin{equation}
\alpha ~=~ \mu + \half \eta ~~,~~ \beta ~=~ 1 - \eta - \chi_\sigma ~~,~~
\gamma ~=~ 1 - \eta - \chi_\pi
\end{equation}
where $\eta$ is the fermion anomalous dimension which is related through the
critical renormalization group equation to the fermion wave function
renormalization constant. The other exponents, $\chi_\sigma$ and $\chi_\pi$,
are the anomalous dimensions of the respective $3$-vertices in (2.1) and these
can also be related to the analogous renormalization constants.

In the critical point approach we discuss throughout this paper, the exponents
are expanded order by order in $1/N$, where $N$ is large, and the exponent
$\eta_i$, say, is calculated at the $i$th order where $\eta$ $=$
$\sum_{i=1}^\infty\eta_i/N^i$ and $\eta_i$ $=$ $\eta_i(\mu,M)$. The method to
do this has been discussed extensively before, \cite{12}, but for completeness
sake and for preparation to calculating the $\beta$-function exponent $\lambda$
we review the calculation of $\eta_1$, \cite{17}. This is achieved by examining
the skeleton Dyson equations with dressed propagators in the critical region,
truncated to the appropriate order in $1/N$, which for the moment will be
leading order. The relevant Feynman graphs for this are illustrated in fig. 1,
where the quantities $\psi^{-1}$, $\sigma^{-1}$ and $\pi^{-1}$ are the
$2$-point functions of the fields. Their asymptotic scaling forms in the
critical region in coordinate space can be easily deduced from the forms of the
propagators (2.4) by mapping (2.4) first to momentum space performing the
inversion there before inverting the map to coordinate space, \cite{5,12}. This
is facilitated by use of the Fourier transform
\begin{equation}
\frac{1}{(x^2)^\alpha} ~=~ \frac{a(\alpha)}{2^{2\alpha}\pi^\mu} \int_k
\frac{e^{ikx}}{(k^2)^{\mu-\alpha}}
\end{equation}
where, for simplicity, $a(\alpha)$ $=$ $\Gamma(\mu-\alpha)/\Gamma(\alpha)$.
The functions are then
\begin{eqnarray}
\psi^{-1}(x) & \sim & \frac{r(\alpha-1)\xslash}{A(x^2)^{2\mu-\alpha+1}}
{}~~,~~ \sigma^{-1}(x) ~ \sim ~ \frac{p(\beta)}{B(x^2)^{2\mu-\beta}}
\nonumber \\
\pi^{-1}(x) & \sim & \frac{p(\gamma)}{C(x^2)^{2\mu-\gamma}}
\end{eqnarray}
as $x^2$ $\rightarrow$ $0$ where
\begin{equation}
p(\beta) ~=~ \frac{a(\beta-\mu)}{\pi^{2\mu}a(\beta)} ~~,~~
r(\alpha-1) ~=~ \frac{\alpha a(\alpha-\mu)}{\pi^{2\mu}(\mu-\alpha)
a(\alpha)}
\end{equation}
To represent the Dyson equations of fig. 1 in the critical region, one merely
substitutes (2.4) in the lines of the one loop graphs to obtain
\begin{eqnarray}
0 &=& r(\alpha-1) + z + 4C_2(R) y \\
0 &=& p(\beta) + 2zNM \\
0 &=& p(\gamma) + 8T(R)Ny
\end{eqnarray}
where $z$ $=$ $A^2B$ and $y$ $=$ $A^2C$ and we have used the relations (2.2)
and (2.3). In (2.9)-(2.11) one has three unknowns, $z$, $y$ and $\eta_1$
through the appearance of $\Gamma(\mu-\alpha)$ in (2.9). Thus eliminating
$z_1$ and $y_1$ after expanding to leading order in $1/N$ one obtains
\begin{equation}
\eta_1 ~=~ \frac{\tilde{\eta}_1}{2} \left[ \frac{1}{M} + \R \right]
\end{equation}
where $\tilde{\eta}_1$ $=$ $-$ $2\Gamma(2\mu-1)/[\Gamma(\mu+1)\Gamma(\mu)
\Gamma(1-\mu)\Gamma(\mu-1)]$ and consequently
\begin{equation}
z_1 ~=~ \frac{\mu\Gamma^2(\mu)\tilde{\eta}_1}{4\pi^{2\mu}M} ~~,~~
y_1 ~=~ \frac{Mz_1}{4T(R)}
\end{equation}
which will be required later. Whilst (2.12) is a result for all $M$ and
agrees in the limits mentioned earlier with \cite{8,9,11} and \cite{12}, it
also checks with a leading order calculation in \cite{11} for $M$ $=$ $2$,
using canonical techniques. However, as has been mentioned this latter
method is not powerful enough to probe to $O(1/N^2)$.

The vertex anomalous dimensions of (2.5) have also been computed to the same
order in $1/N$ by examining the scaling behaviour of the $3$-point functions
in an analogous way at criticality by extending the earlier work of
\cite{24,25} and we record the results in preparation for computing
$\lambda$, as
\begin{eqnarray}
\chi_{\sigma \, 1} &=& \frac{\mu\tilde{\eta}_1}{2(\mu-1)} \left[ \M - \R
\right] \\
\chi_{\pi \, 1} &=& \frac{\mu\tilde{\eta}_1}{2(\mu-1)} \left[ \R - \M
- \frac{C_2(G)}{2T(R)} \right]
\end{eqnarray}
Again each agrees in the appropriate limit with earlier results and for
$M$ $=$ $2$, $\chi_{\sigma \, 1}$ agrees with \cite{11}.

\sect{$\beta$-function exponent.}
Having indicated the simplicity of the method to deduce the exponents of all
the fields at leading order, we now present the extensions to the formalism to
deduce $\lambda_1$ which will serve as the foundation for calculating
$\lambda_2$ where $\lambda$ $=$ $\mu$ $-$ $1$ $+$ $\sum_{i=1}^\infty
\lambda_i/N^i$. To achieve this one considers the corrections to the
asymptotic scaling forms (2.4) by including $O(x^2)$ terms, \cite{5,6,12},
\begin{eqnarray}
\psi(x) & \sim & \frac{A\xslash}{(x^2)^\alpha} [ 1+A^\prime(x^2)^\lambda ] \\
\sigma(x) & \sim & \frac{B}{(x^2)^\beta} [ 1+B^\prime(x^2)^\lambda ] \\
\pi(x) & \sim & \frac{C}{(x^2)^\gamma} [ 1+C^\prime(x^2)^\lambda ]
\end{eqnarray}
Additional higher order corrections, which will involve other exponents such as
the specific heat, could also be computed by including the appropriate forms.
However, as there exist hyperscaling laws relating such exponents, by doing
this one would merely succeed in demonstrating their consistency which is not
in doubt, \cite{5,6}. The quantities $A^\prime$, $B^\prime$ and $C^\prime$ are
the respective amplitudes associated with each correction and the analogous
$2$-point functions can be deduced in a similar way to (2.7) to obtain,
\cite{12},
\begin{eqnarray}
\psi^{-1}(x) &\sim& \frac{r(\alpha-1)\xslash}{A(x^2)^{2\mu-\alpha+1}}
[1-A^\prime s(\alpha-1)(x^2)^\lambda] \\
\sigma^{-1}(x) &\sim& \frac{p(\beta)}{B(x^2)^{2\mu-\beta}}
[1-B^\prime q(\beta)(x^2)^\lambda] \\
\pi^{-1}(x) &\sim& \frac{p(\gamma)}{C(x^2)^{2\mu-\gamma}}
[1-C^\prime q(\gamma)(x^2)^\lambda]
\end{eqnarray}
as $x^2$ $\rightarrow$ $0$ where
\begin{equation}
q(\alpha) ~=~ \frac{a(\alpha-\mu+\lambda)a(\alpha-\lambda)}
{a(\alpha-\mu)a(\alpha)} ~~,~~ s(\alpha) ~=~ \frac{q(\alpha) \alpha
(\alpha-\mu)}{(\alpha-\mu+\lambda)(\alpha-\lambda)}
\end{equation}

With these corrections one can now reconsider the Dyson equations in the
critical region. However, when one is dealing with a theory where the
fundamental field is fermionic in this large $N$ approach, as we have here,
several higher order graphs have to be included in the $\sigma$ and $\pi$ Dyson
equations, \cite{22}, which are illustrated in fig. 2. The reason for this is
quite simple and can be seen, for instance, by examining the $N$-dependence of
each term in the $\sigma$-consistency equation. First, we have from figs 1 and
2, in the critical region
\begin{eqnarray}
0 &=& \frac{p(\beta)}{N}[1-B^\prime q(\beta)(x^2)^\lambda]
+ 2zM[1+2A^\prime(x^2)^\lambda] \nonumber \\
&-& z^2M[\Pi + (x^2)^\lambda(A^\prime\Pi_{1A}+B^\prime\Pi_{1B})] \nonumber \\
&+& 4yzMC_2(R)[\Pi + (x^2)^\lambda(A^\prime\Pi_{1A} + C^\prime\Pi_{1C})]
\end{eqnarray}
where the subscripts $A$, $B$ and $C$ denote the respective insertion of
$(x^2)^\lambda$ on a $\psi$, $\sigma$ or $\pi$ line. As in \cite{6,12} the
equation decouples into two pieces one of which is relevant for the computation
of $\eta_2$, which we ignore here, and the other which involves the
$(x^2)^\lambda$ terms, since each is of differing dimension in $x^2$, ie
\begin{equation}
0 ~=~ 4zA^\prime - B^\prime\left[\frac{p(\beta)q(\beta)}{NM}
+ z^2\Pi_{1B}\right] + 4C_2(R)yz\Pi_{1C}
\end{equation}
In (3.9), we have neglected the graphs where there is an insertion on the
fermion line of the $2$-loop graphs of fig. 2 as these do not contribute at
$O(1/N)$. Further, the expressions $\Pi_{1A}$, $\Pi_{1B}$ and $\Pi_{1C}$ denote
the value of the integral which is $O(1)$, without symmetry factors. Now if one
substitutes the leading order values for $\alpha$, $\beta$ and $\gamma$ into
(3.9) to examine the location of $N$ in each term, it is easy to see that both
terms of the coefficient of $B^\prime$ are of the same order. Therefore,
neglecting the graph $\Pi_{1B}$ would omit a contribution and lead to an
erroneous result and so it must be included. Similarly, the $\pi$ consistency
equation gives
\begin{eqnarray}
0 &=& 16T(R)yA^\prime + 4T(R)yz\Lambda_{1B}B^\prime \nonumber \\
&-& C^\prime\left[ \frac{p(\gamma)q(\gamma)}{N} + 8T(R)y^2 (2C_2(R)-C_2(G))
\Lambda_{1C} \right]
\end{eqnarray}

Finally, in the $\psi$ equation one needs only to consider the one loop graphs
of fig. 1 since there is no reordering as there is in (3.9) and (3.10) which
can be seen by inspection. Thus
\begin{equation}
0 ~=~ A^\prime[-r(\alpha-1)s(\alpha-1) + z + 4yC_2(R)] + zB^\prime
+ 4yC_2(R)C^\prime
\end{equation}
To proceed one forms a $3$ $\times$ $3$ matrix with $A^\prime$, $B^\prime$
and $C^\prime$ as the basis vectors and sets its determinant to zero for
consistency whence an expression for $\lambda_1$ will emerge since it is
the only unknown. For the moment the matrix is
\begin{equation}
\left(
\begin{array}{ccc}
r(\alpha-1)s(\alpha-1) & z & C_2(R)y \\
4z & \frac{p(\beta)q(\beta)}{NM} + z^2\Pi^\prime & - \, C_2(R)yz\Pi^\prime \\
4y & - \, yz\Pi^\prime & \frac{p(\gamma)q(\gamma)}{16T(R)N}
+ \frac{y^2\Pi^\prime (2C_2(R)-C_2(G))}{2} \\
\end{array}
\right)
\end{equation}
where the explicit calculation of the $2$-loop graphs gives $\Pi_{1B}$ $=$
$\Pi_{1C}$ $=$ $\Lambda_{1B}$ $=$ $\Lambda_{1C}$ $\equiv$ $\Pi^\prime$ $=$
$2\pi^{2\mu}/ [\Gamma^2(\mu)(\mu-1)^2]$, \cite{12,22}. Again it is easy to see
by examining the $N$-dependence in each element of (3.12) the necessity of
including the graphs of fig. 2. To solve for $\lambda_1$ by setting the
determinant of (3.12) to zero one makes several row and column transformations.
Subtracting row two from row three and then column three from column two one
obtains the following matrix whose determinant is set to zero
\begin{equation}
\left(
\begin{array}{ccc}
r(\alpha-1)s(\alpha-1) & z\left(1 + \frac{MC_2(R)}{T(R)}\right)
& \frac{MC_2(R)z}{2T(R)} \\
- \, 4 & \! 2q(\beta) - \Pi^\prime \! \left( 1 - \frac{MC_2(R)}{T(R)}\right) &
\frac{MC_2(R)z\Pi^\prime}{2T(R)} \\
0 & 0 & \!\!\! q(\gamma) - \frac{Mz\Pi^\prime(4C_2(R)-C_2(G))}{4T(R)} \\
\end{array}
\right)
\end{equation}
where we have used (2.13). Thus setting (3.12) to zero yields $\lambda_1$ as
\begin{equation}
\lambda_1 ~=~ - \, \frac{(2\mu-1)\tilde{\eta}_1}{2} \left[ \M + \R \right]
\end{equation}
This agrees with previous results of the $O(N)$ model \cite{12} and the
$U(1)$ $\times$ $U(1)$ case \cite{11,26} in the appropriate limits as well as
the $SU(2)$ $\times$ $SU(2)$ calculation of \cite{11}. However, it is worth
noting that an alternative second solution might appear to emerge from the
lower right element of (3.13). We discard this as it is not consistent with
previous results and, moreover, the row and column transformations we have
made do not induce any contribution from the $\psi$ equations which appear
as the upper row in (3.12). Indeed this type of factorization of the final
determinant has been observed in other contexts, \cite{22}, where one had also
to ignore a spurious solution. Also, we remark that together with (2.12),
(2.14) and (2.15) we now have a complete set of exponents for the $SU(M)$
$\times$ $SU(M)$ model at leading order.

We conclude this section by stating that it is now possible to proceed
beyond (3.14) and attempt to calculate the $O(1/N^2)$ corrections, which has
been achieved recently for the $O(N)$ model \cite{14,15} and $U(1)$
$\times$ $U(1)$ case \cite{26} and use can be made of results calculated
there. Something which is necessary for this is the $O(1/N^2)$ corrections
to $\eta$, $\chi_\sigma$ and $\chi_\pi$ since these will appear in the
$1/N$ expansion of the functions of (3.4)-(3.6). They have been deduced in
\cite{17} by considering the scaling behaviour of the higher order graphs in
the Dyson equations of the $2$ and $3$ point functions with dressed
propagators and we record the results here. We have, \cite{17},
\begin{eqnarray}
\eta_2 &=& \frac{\tilde{\eta}^2_1}{4} \left[ \left(\M + \R \right)^2
\left(\Psi(\mu) + \frac{2}{\mu-1} + \frac{1}{2\mu} \right) \right. \\
&+& \left. \frac{\mu}{(\mu-1)} \left( \left( \M-\R\right)^2
- \frac{C_2(G)C_2(R)}{2T^2(R)} \right)
\left( \Psi(\mu) + \frac{3}{2(\mu-1)} \right) \right] \nonumber
\end{eqnarray}
(the sign of the coefficient of the $C_2(G)$ term was incorrectly given in
\cite{17})
\begin{eqnarray}
\chi_{\sigma \, 2} &=& \frac{\mu\tilde{\eta}^2_1}{4(\mu-1)^2}
\left[ (2\mu-1)\left( \frac{1}{M^2} - \frac{C^2_2(R)}{T^2(R)} \right)
\left( \Psi(\mu) + \frac{1}{(\mu-1)}\right) \right. \nonumber \\
&+& \left. \frac{\mu C_2(R)C_2(G)}{2T^2(R)} \left( \Psi(\mu)
+ \frac{1}{(\mu-1)} \right) + \frac{3\mu}{2(\mu-1)} \left( \frac{1}{M}
- \frac{C_2(R)}{T(R)}\right)^2 \right. \nonumber \\
&+& \left. \frac{5\mu C_2(R)}{(\mu-1)T(R)} \left(\M - \R\right)
- \frac{2\mu}{(\mu-1)}\left(\frac{1}{M^2} - \frac{C^2_2(R)}{T^2(R)} \right)
\right. \nonumber \\
&+& \left. \frac{\mu}{2(\mu-1)}\left(\M+\R\right)^2
- \frac{\mu(2\mu^2-5\mu+4)}{(\mu-1)M} \left(\M + \frac{3C_2(R)}{T(R)} \right)
\right. \nonumber \\
&+& \left. \frac{\mu}{M}\left( 3(\mu-1)\Theta(\mu)
- \frac{(2\mu-3)}{(\mu-1)}\right) \left(\M - \R\right) \right]
\end{eqnarray}
and
\begin{eqnarray}
\chi_{\pi \, 2} &=& \frac{\mu\tilde{\eta}^2_1}{4(\mu-1)^2}
\left[ \Psi(\mu) + \frac{1}{(\mu-1)} \right] \nonumber \\
&& \times \left[(2\mu-1)\left(\M+\R\right)
\left(\R-\M-\frac{C_2(G)}{2T(R)}\right) \right. \nonumber \\
&&- \left. \frac{\mu C_2(G)}{2T(R)} \left(\R-\frac{2}{M}-\frac{C_2(G)}{2T(R)}
\right) \right] \nonumber \\
&+& \frac{3\mu^2\tilde{\eta}^2_1}{4(\mu-1)^3}
\left(\R-\M-\frac{C_2(G)}{2T(R)}\right)^2 \nonumber \\
&+& \frac{5\mu^2\tilde{\eta}^2_1}{16(\mu-1)^3M} \left[ 4 \left(\R-\M\right)
- \G \right] \nonumber \\
&+& \frac{2\mu^2\tilde{\eta}^2_1}{(\mu-1)^3} \left[ \frac{1}{4} \left(
\frac{1}{M^2} - \frac{C_2(G)}{2T(R)M}
- \left(\R - \frac{C_2(G)}{2T(R)}\right)^2 \right) \right. \nonumber \\
&+& \left. \frac{1}{16} \left( \left( \R+\M\right)^2 - \frac{C_2(G)}
{MT(R)} - \frac{3C_2(G)C_2(R)}{2T^2(R)} + \frac{C^2_2(G)}{2T^2(R)} \right)
\right. \nonumber \\
&-& \left. \frac{3}{16}\left( \M - \R + \frac{C_2(G)}{2T(R)} \right)^2
- \frac{(2\mu^2-5\mu+4)}{8} \left( \frac{3}{M^2}
+ \frac{C_2(R)}{T(R)M} \right. \right. \nonumber \\
&-& \left. \left. \frac{C_2(G)}{4T(R)}\left(\R-\frac{3}{M}\right)
+\left(\R-\M\right)^2 \right) \right. \nonumber \\
&+& \left. \frac{(\mu-1)^2}{8} \left( 3\Theta(\mu) - \frac{(2\mu-3)}{(\mu-1)^2}
\right) \right. \nonumber \\
&& \left. \times \left( \R - \M - \frac{C_2(G)}{2T(R)} \right)
\left( \R - \frac{C_2(G)}{4T(R)} \right) \right]
\end{eqnarray}
where $\Psi(\mu)$ $=$ $\psi(2\mu-1)$ $-$ $\psi(1)$ $+$ $\psi(2-\mu)$ $-$
$\psi(\mu)$ and $\Theta(\mu)$ $=$ $\psi^\prime(\mu)$ $-$ $\psi^\prime(1)$ where
$\psi(x)$ is the logarithmic derivative of the $\Gamma$-function. Further, from
the $\eta_2$ consistency equations
\begin{eqnarray}
z_2 &=& \frac{\mu\Gamma^2(\mu)\tilde{\eta}^2_1}{8\pi^{2\mu}M}
\left[ \left( \M + \R + \frac{\mu}{(\mu-1)} \left( \M - \R \right)
\right) \right. \nonumber \\
&& \left. \times \left( \Psi + \frac{2}{(\mu-1)} \right)
- \frac{\mu}{(\mu-1)^2} \left( \M - \R \right) \right] \\
y_2 &=& \frac{\mu\Gamma^2(\mu)\tilde{\eta}^2_1}{32\pi^{2\mu}T(R)}
\left[ \left( \M + \R + \frac{\mu}{(\mu-1)} \left( \R - \M - \frac{C_2(G)}
{2T(R)} \right) \right) \right. \nonumber \\
&& \left. \times \left( \Psi + \frac{2}{(\mu-1)} \right)
- \frac{\mu}{(\mu-1)^2} \left( \R - \M - \frac{C_2(G)}{2T(R)} \right) \right]
\end{eqnarray}

\sect{Beyond $O(1/N)$.}
The main effort required to go beyond (3.14) and determine $\lambda_2$ involves
computing the higher order Feynman graphs which appear in the Dyson equations
of each field. First, though we need to derive the formal corrections to each
critical representation of the Dyson equations which is a non-trivial
exercise due to the appearance of divergent graphs which have a vertex
subgraph and finite three and four loop graphs. The renormalization
techniques required to handle the infinities have been discussed in other
places, \cite{6,12}, but for completeness we detail the procedure here and
concentrate on the $\psi$ equation since there are fewer corrections to be
considered compared to the $\sigma$ and $\pi$ fields because of the extra
graphs which arise. The higher order graphs with dressed propagators for
the $\psi$ Dyson equations are illustrated in fig. 3. Including them in (2.9)
and (3.11) one obtains
\begin{eqnarray}
0 &=& r(\alpha-1)[1-A^\prime s(\alpha-1)(x^2)^\lambda]
+ z [1+(A^\prime+B^\prime)(x^2)^\lambda]u^2(x^2)^{\chi_\sigma+\Delta}
\nonumber \\
&+& 4C_2(R)y[1+(A^\prime+C^\prime)(x^2)^\lambda]v^2(x^2)^{\chi_\pi+\Delta}
\nonumber \\
&+& z^2[\Sigma+(A^\prime\Sigma_A + B^\prime\Sigma_B)(x^2)^\lambda]
(x^2)^{2\chi_\sigma+2\Delta} \\
&-& 8yzC_2(R)[\Sigma+(A^\prime\Sigma_A+B^\prime\Sigma_B+C^\prime\Sigma_C)
(x^2)^\lambda](x^2)^{\chi_\sigma+\chi_\pi+2\Delta} \nonumber \\
&+& 16y^2C_2(R)[C_2(R) - \half C_2(G)][\Sigma+(A^\prime\Sigma_A
+C^\prime\Sigma_C)(x^2)^\lambda](x^2)^{2\chi_\pi+2\Delta} \nonumber
\end{eqnarray}
Unlike at leading order the powers of $x^2$ do not cancel here and, moreover,
the graphs $\Sigma$, $\Sigma_A$, $\Sigma_B$ and $\Sigma_C$ are divergent
with each having simple poles in $\Delta$. This infinitesimal quantity is a
regularization introduced to control such infinities by shifting the
exponents of $\sigma$ and $\pi$ by $\beta$ $\rightarrow$ $\beta$ $-$
$\Delta$, $\gamma$ $\rightarrow$ $\gamma$ $-$ $\Delta$, \cite{6,12}. If we
formally define the $\Delta$-finite part of the two loop corrections to be
$\Sigma^\prime_A$ etc the simple poles in the $(x^2)^\lambda$ sector of (4.1)
are absorbed by choosing the respective vertex counterterms $u$ and $v$ in
such a way that (4.1) is $\Delta$-finite at $O(1/N^2)$. After this the
$\ln x^2$ terms which remain and which prevent one approaching the $x^2$
$\rightarrow$ $0$ region are removed by choosing $\chi_{\sigma \, 1}$ and
$\chi_{\pi \, 1}$ appropriately. It turns out that the choices already found in
the analysis of the $3$-point functions, (2.14) and (2.15), achieve this
which is a consistency check on our renormalization. Decoupling the equation
as before yields
\begin{eqnarray}
0 &=& A^\prime[-r(\alpha-1)s(\alpha-1)+z+4yC_2(R) \nonumber \\
&+& \Sigma^\prime_A(z^2 - 8yzC_2(R) + 16y^2(C_2(R)-\half C_2(G)))] \nonumber \\
&+& B^\prime[z + z\Sigma^\prime_B(z-8yC_2(R))] \nonumber \\
&+& C^\prime[4C_2(R)y + 8yC_2(R)\Sigma_C^\prime(y(2C_2(R)-C_2(G))-z)]
\end{eqnarray}
which is the $O(1/N^2)$ correction to (3.11). However, examining the
$N$-dependence of the terms in the coefficient of $A^\prime$,
$\Sigma^\prime_A$ is $O(1/N^2)$ relative to $r(\alpha-1)s(\alpha-1)$ and can
therefore be neglected in the overall $3$ $\times$ $3$ determinant.
Moreover, the explicit calculation of $\Sigma^\prime_B$ and $\Sigma^\prime_C$
reveals that both are zero \cite{12,15} and therefore (4.2) simplifies to
\begin{equation}
0 ~=~ A^\prime[-r(\alpha-1)s(\alpha-1) + z + 4yC_2(R)]
+ z B^\prime + 4C_2(R)y C^\prime
\end{equation}
where of course now the $O(1/N^2)$ terms of $z$ and $y$ are needed. We have
detailed the renormalization of the $\psi$ equation since it has a simpler
structure than that of the $\sigma$ and $\pi$ equations which are complicated
by the higher order graphs which have to be included. The relevant graphs, with
dressed propagators, for the $\sigma$ equation are illustrated in fig. 4 where
the label beside each graph will be used in the following. We have only
displayed the distinct topologies in fig. 4 and the solid internal lines,
without dots, denote either a $\sigma$ or $\pi$ field. As at leading order we
need only consider $(x^2)^\lambda$ insertions on the bosonic lines, since
within the final $3$ $\times$ $3$ matrix graphs with insertions on the
fermionic lines will contribute to $\lambda_3$ only. The dot is intended to
indicate the location of the $(x^2)^\lambda$ insertion when one substitutes the
asymptotic scaling forms of the propagators (3.1)-(3.3) in the graphs. The
correction graphs for the $\pi$ equations are formally equivalent to those of
fig. 4. It is clear from the structure of $\Pi_{2B1}$ and $\Pi_{3B}$ that they
are $\Delta$-divergent due to the presence of a vertex subgraph. Within the
consistency equation the same renormalization procedure and vertex counterterms
we discussed for the $\psi$ equation removes the simple poles in $\Delta$ and
it is therefore a straightforward matter to write down the finite $O(1/N^2)$
correction to (3.9) in the critical region as
\begin{eqnarray}
0 &=& zMA^\prime[4-z\Pi_{1A}] - B^\prime\left[ \frac{p(\beta)q(\beta)}{N}
+ z^2M\Pi_{1B} + \Pi_{B2} \right] \nonumber \\
&+& C^\prime[4C_2(R)Myz\Pi_{1C} - \Pi_{C2} ]
\end{eqnarray}
where $\Pi_{B2}$ and $\Pi_{C2}$ are formally defined to be
\begin{eqnarray}
\Pi_{B2} &=& 2(\Pi_{2B1} + \Pi_{2B2}) + \Pi_{3B} + \Pi_{4B} \nonumber \\
&-& 2(2\Pi_{5B1} + \Pi_{5B2}) - 4(\Pi_{6B1} + 2\Pi_{6B2}) \\
\Pi_{C2} &=& 2(\Pi_{2C1} + \Pi_{2C2}) + \Pi_{3C} + \Pi_{4C} \nonumber \\
&-& 2(2\Pi_{5C1} + \Pi_{5C2}) - 4(\Pi_{6C1} + 2\Pi_{6C2})
\end{eqnarray}
with the $\Delta$-finite part only understood as contributing. The factors of
$z$ and $y$ as well as group factors are contained within each formal term of
(4.5) and (4.6) and will be computed explicitly later. Whilst $\Pi_{1B}$ and
$\Pi_{1C}$ contributed at leading order it is important to remember that each
has an $O(1/N)$ correction which will appear in $\lambda_2$. Further, explicit
calculation has shown that $\Pi^\prime_{1A}$ $=$ $0$, \cite{12}. The formal
corrections to the $\pi$ equation are also now straightforward to write down
and we find
\begin{eqnarray}
0 &=& 16T(R)yA^\prime + B^\prime[4T(R)yz\Lambda_{1B} - \Lambda_{B2}]
\nonumber \\
&-& C^\prime\left[ \frac{p(\gamma)q(\gamma)}{N}
+ 8y^2T(R)(2C_2(R)-C_2(G))\Lambda_{1C} + \Lambda_{C2} \right]
\end{eqnarray}
where the higher order corrections, $\Lambda_{B2}$ and $\Lambda_{C2}$, have the
same formal definitions as (4.5) and (4.6). Thus, with (4.2), (4.4) and (4.7)
one can now form the corrected version of the $3$ $\times$ $3$ matrix (3.12)
and deduce the master equation which yields $\lambda_2$. All that remains is
the explicit evaluation of the higher order graphs.

These fall into two classes. First, we need to consider the $O(1/N)$
corrections to $\Pi_{1B}$, $\Pi_{1C}$, $\Lambda_{1B}$ and $\Lambda_{1C}$ since,
for example, the fermionic lines of the $2$-loop graphs each have exponents
$\mu$ $+$ $\half\eta$ which therefore means they have to be expanded in powers
of $1/N$. An algorithm to do this was presented in \cite{27} and the analogous
graph has also been computed in the $O(N)$ model, \cite{12,22}. Whilst the
graphs are of a similar structure the exponents of the internal bosonic lines
differ in each case and we therefore had to recompute $\Pi_{1B}$ $=$ $\Pi_{1C}$
and $\Lambda_{1B}$ $=$ $\Lambda_{1C}$ explicitly. Using the techniques of
\cite{27} we obtained
\begin{eqnarray}
\Pi_{1B} &=& \frac{2\pi^{2\mu}}{(\mu-1)^2\Gamma^2(\mu)} \left[
1 - \frac{\tilde{\eta}_1}{(\mu-1)N} \left( \M + \R \right) \right. \nonumber \\
&+& \left. \frac{3\mu(\mu-1)\tilde{\eta}_1}{2N}
\left( \Theta(\mu) + \frac{1}{(\mu-1)^2} \right) \right. \nonumber \\
&& \times ~ \left. \left( \M + \R - \frac{1}{2(\mu-1)}
\left(\M-\R\right)\right)\right] \\
\Lambda_{1B} &=& \frac{2\pi^{2\mu}}{(\mu-1)^2\Gamma^2(\mu)} \left[ 1
- \frac{\tilde{\eta}_1}{(\mu-1)N} \left( \M + \R \right) \right. \nonumber \\
&+& \left. \frac{3\mu(\mu-1)\tilde{\eta}_1}{2N} \left( \Theta(\mu) +
\frac{1}{(\mu-1)^2} \right) \right. \nonumber \\
&& \times ~ \left. \left( \M + \R - \frac{1}{2(\mu-1)}
\left(\R-\M - \frac{C_2(G)}{2T(R)}\right)\right)\right] \nonumber \\
\end{eqnarray}

The second class of graphs are those of fig. 4. Since the integral structure
of each is completely the same as the graphs of the $O(N)$ model we needed
only to compute the $SU(M)$ factors multiplying each which arise from the
appearance of $\lambda^a\gamma^5$ at various vertices of the graphs. Making
use of the relations, which have been discussed in \cite{28},
\begin{equation}
\lambda^a \lambda^b ~=~ \frac{4T(R)}{M} \delta^{ab} I + d^{abc} \lambda^c
+ i f^{abc} \lambda^c
\end{equation}
which defines the totally symmetric tensor $d^{abc}$ in $SU(M)$ and
\begin{eqnarray}
d^{apq} d^{bpq} &=& \left[ 4C_2(R) - \frac{4T(R)}{M} - C_2(G) \right]
\delta^{ab} \\
d^{apq}f^{brp}f^{crq} &=& \frac{C_2(G)}{2} d^{abc} \\
d^{apq}d^{brp}d^{crq} &=& \left[ 4C_2(R) - \frac{8T(R)}{M}
- \frac{3C_2(G)}{2} \right] d^{abc}
\end{eqnarray}
we managed to compute $\Pi_{B2}$, $\Pi_{C2}$, $\Lambda_{B2}$ and $\Lambda_{C2}$
relative to the $O(N)$ model graphs, which are denoted by a superscript
${}^{\mbox{o}}$ in the following. Making use of the relation between $y_1$
and $z_1$ we have
\begin{eqnarray}
\Pi_{B2} &=& 2z^3 M^2 \left( \M + \R \right) \left[ \Pio_{2B2}
+ \Pio_{4B} \right. \nonumber \\
&& -~ \left. z_1 M ( 2\Pio_{5B1} + \Pio_{5B2} + 2\Pio_{6B1)} \right]
\nonumber \\
&+& 2z^3M^2 \left( \M - \R \right) \left[ \Pio_{2B1} + \Pio_{3B}
- 4 z_1 M \Pio_{6B2} \right]
\end{eqnarray}
\begin{eqnarray}
\Pi_{C2} &=& \frac{2z^3M^3C_2(R)}{T(R)} \left[ \left( \M - \R
+ \frac{C_2(G)}{2T(R)}\right) \Pio_{2B1} \right. \nonumber \\
&-& \left. \left( \M + \R - \frac{C_2(G)}{2T(R)} \right) \Pio_{2B2}
- \left( \M - \R \right) \Pio_{3B} \right. \\
&+& \left. \left( \M + \R - \frac{C_2(G)}{2T(R)} \right) \Pio_{4B}
- 2z_1 (2\Pio_{5B1} + \Pio_{5B2} - 2 \Pio_{6B1}) \frac{}{} \right] \nonumber
\end{eqnarray}
\begin{eqnarray}
\Lambda_{B2} &=& - \, 2M^2z^3 \left[ \left( \M + \R - \frac{C_2(G)}{2T(R)}
\right) \Pio_{2B2} \right. \nonumber \\
&+& \left. \left( \M - \R \right) \Pio_{2B1}
- \left(\M - \R + \frac{C_2(G)}{2T(R)} \right) \Pio_{3B} \right. \\
&-& \left. \left( \M + \R - \frac{C_2(G)}{2T(R)} \right) \Pio_{4B}
+ 2z_1 (2\Pio_{5B1} + \Pio_{5B2} - 2 \Pio_{6B1}) \frac{}{}\right] \nonumber
\end{eqnarray}
\begin{eqnarray}
\Lambda_{C2} &=& 2z^3M^3 \left[ \left(\R - \frac{C_2(G)}{2T(R)}\right)
\left(\R-\M-\frac{C_2(G)}{2T(R)}\right)\Pio_{2B1} \right. \nonumber \\
&+& \left. \left( \frac{C_2(R)}{MT(R)} + \left( \R - \frac{C_2(G)}{2T(R)}
\right)^2 \right) \Pio_{2B2} \right. \nonumber \\
&+& \left. \left( \R - \frac{C_2(G)}{2T(R)}\right)\left( \R - \M
- \frac{C_2(G)}{2T(R)}\right) \Pio_{3B} \right. \nonumber \\
&+& \left. \left( \R - \frac{C_2(G)}{2T(R)}\right)\left( \M + \R
- \frac{C_2(G)}{2T(R)}\right) \Pio_{4B} \right. \nonumber \\
&-& \left. z_1M \! \left( \! \left( \frac{1}{M^2} + \frac{C_2(R)}{MT(R)}
+ \left( \R - \M - \frac{C_2(G)}{4T(R)}\right)^2\!\! +
\frac{C_2^2(G)}{16T^2(R)}
\right) \right. \right. \nonumber \\
&& \left. \left. \times~ (2\Pio_{5B1} + \Pio_{5B2} ) \right. \right.
\nonumber \\
&+& \left. \left. 2\left( \M \left( \M + \R - \frac{C_2(G)}{2T(R)}\right)
\right. \right. \right. \nonumber \\
&& +~ \left. \left. \left.  \left( \R - \M - \frac{C_2(G)}{4T(R)}
\right) \left( \R - \M - \frac{C_2(G)}{2T(R)} \right) \right) \Pio_{6B1}
\right. \right. \nonumber \\
&+& \left. \left. 4\left( \R - \M - \frac{C_2(G)}{2T(R)} \right)
\left( \R - \frac{C_2(G)}{4T(R)} \right) \Pio_{6B2} \right) \right]
\end{eqnarray}
The explicit values of the basic graphs of the $O(N)$ model are, \cite{15},
\begin{eqnarray}
\Pio_{2B1} &=& - \, \frac{2\pi^{4\mu}}{(\mu-1)^3\Gamma^4(\mu)\Delta}
\left[ 1 - \Delta (\mu-1) \left( 3\Theta + \frac{2}{(\mu-1)^2} \right) \right]
\\
\Pio_{2B2} &=& - \, \frac{2\pi^{4\mu}}{(\mu-1)^4\Gamma^4(\mu)} \\
\Pio_{3B~} &=& - \, \frac{2\pi^{4\mu}}{(\mu-1)^3\Gamma^4(\mu)\Delta}
\left[ 1 - \frac{\Delta(\mu-1)}{2} \left( 3\Theta + \frac{1}{(\mu-1)^2}
\right)\right] \\
\Pio_{4B~} &=& \frac{\pi^{4\mu}}{(\mu-1)^2\Gamma^4(\mu)}
\left[ 3\Theta + \frac{1}{(\mu-1)^2} \right] \\
\Pio_{5B1} &=& \frac{(2\mu-3)\pi^{6\mu}a(2\mu-2)}{2(\mu-1)^5(\mu-2)
\Gamma^3(\mu)} \left[ 6\Theta - \Phi - \Psi^2 + \frac{5}{2(\mu-1)^2}
- \frac{8}{(2\mu-3)} \right. \nonumber \\
&+& \left. \frac{1}{(2\mu-3)(\mu-1)} + \frac{\Psi}{(2\mu-3)(\mu-1)}
+ \frac{2(\mu-2)\Psi}{(\mu-1)} + \frac{(\mu-2)}{(\mu-1)^2} \right] \nonumber \\
\Pio_{5B2} &=& - \, \frac{\pi^{6\mu}a(2\mu-2)}{(\mu-1)^5\Gamma^3(\mu)}
\left[ \frac{(2\mu-3)}{(\mu-2)}\left( \Phi + \Psi^2 - \frac{1}{2(\mu-1)^2}
\right) \right. \nonumber \\
&-& \left. \frac{(3\mu-4)\Psi}{(\mu-1)(\mu-2)^2} + \frac{1}{(\mu-2)^2}
\right] + \frac{2\pi^{6\mu}a^2(2\mu-2)}{(\mu-1)^6(\mu-2)^2} \\
\Pio_{6B1} &=& \frac{\pi^{6\mu}a(2\mu-2)}{(\mu-1)^5\Gamma^3(\mu)}
\left[ \frac{1}{(\mu-1)} - \frac{5}{2(\mu-1)^2} - \frac{(2\mu-1)\Psi}{(\mu-1)}
\right] \nonumber \\
&+& \frac{\pi^{6\mu}a^2(2\mu-2)}{(\mu-1)^8} \\
\Pio_{6B2} &=& - \, \frac{(2\mu-3)\pi^{6\mu}a(2\mu-2)}{(\mu-1)^5(\mu-2)}
\left[ \frac{\Phi}{2} + \frac{\Psi^2}{2} - \frac{3(\mu-1)}{(2\mu-3)}
\left( \Theta + \frac{1}{(\mu-1)^2} \right) \right. \nonumber \\
&+& \left. \frac{\Psi}{2(\mu-1)} - \frac{1}{4(\mu-1)^2}
+ \frac{(\mu-2)\Psi}{(2\mu-3)(\mu-1)} \right. \nonumber \\
&+& \left. \frac{(\mu-2)}{2(2\mu-3)(\mu-1)^2} + \frac{2}{(2\mu-3)(\mu-1)}
\right] - \frac{\pi^{6\mu}a^2(2\mu-2)}{(\mu-1)^7(\mu-2)}
\end{eqnarray}
where $\Phi(\mu)$ $=$ $\psi^\prime(2\mu-1)$ $-$ $\psi^\prime(2-\mu)$ $-$
$\psi^\prime(\mu)$ $+$ $\psi^\prime(1)$. It is worth recalling that in the
$U(1)$ $\times$ $U(1)$ case, \cite{26}, only three of the above were
relevant for $\lambda_2$ due to the high degree of symmetry present in that
model.

\sect{Discussion.}
We are now in a position to substitute for all the quantities in the overall
$3$ $\times$ $3$ matrix and set its determinant to zero. If one follows the
same formal transformations which were made at leading order to simplify the
evaluation we notice that the factorization of the matrix into two solutions,
one of which is irrelevant, does not occur at $O(1/N^2)$. For instance, the
$(13)$ element is, up to factors,
\begin{eqnarray}
4T(R)y - Mz &=& \frac{\mu^2\Gamma^2(\mu)\tilde{\eta}_1}{8(\mu-1)\pi^{2\mu}
N^2} \left[ \frac{2C_2(R)}{T(R)} - \frac{2}{M} - \frac{C_2(G)}{2T(R)}
\right] \nonumber \\
&& \times ~ \left[ \Psi(\mu) + \frac{1}{(\mu-1)} \right]
\end{eqnarray}
which is zero at leading order but not at $O(1/N^2)$. Similarly the $(23)$
element also vanishes at $O(1/N)$ but not at the subsequent order for all $M$.
This is a rather unfortunate situation especially given the fact that it was
possible to write down $O(1/N^2)$ expressions for $\eta$, $\chi_\sigma$ and
$\chi_\pi$ and $\lambda$ at $O(1/N)$ for all $M$ for this model. The absence of
a solution can, however, be related to what we will term as the non-realisation
of the chiral symmetry of the model. For instance, it is easy to see that when
$M$ $=$ $2$ in the complicated expressions (3.16) and (3.17), $\chi_\sigma$ $=$
$\chi_\pi$ to $O(1/N^2)$, but this simple relation does not hold for $M$ $>$
$2$, \cite{29}. In the $U(1)$ $\times$ $U(1)$ model, \cite{17}, $\chi_\sigma$
is also equivalent to $\chi_\pi$ at $O(1/N^2)$ which indicated that the chiral
symmetry is preserved in that instance. Now if one examines (5.1) it is zero at
$M$ $=$ $2$ and no other (positive) value of $M$ for all $\mu$ since the group
factor is proportional to $(M^2-4)$. Whilst the different behaviour of the $M$
$=$ $2$ and $M$ $>$ $2$ models has been suggested in \cite{11} where the
non-realisation of the chiral symmetry was postulated this is as far as we are
aware the first observation of it in an explicit exponent calculation.

We now restrict attention to $M$ $=$ $2$ as this is the only case we can obtain
$\lambda_2$ in all dimensions. (Of course, it would be possible to derive an
expression for $\lambda_2$ for $M$ $>$ $2$ as the solution of a quadratic
equation, but the complicated and non-linear nature of such an exponent would
not be in keeping with the structure of exponents found in other models. Indeed
a similar situation has also arisen before in \cite{22} when the
non-factorization of the consistency determinant indicated that that exponent
was not a sensible quantity to compute.) It is worth recording the explicit
values of the various exponents when $M$ $=$ $2$ in order to solve (4.2), (4.4)
and (4.7) as, \cite{17},
\begin{eqnarray}
\eta_1 &=& \tilde{\eta}_1 \\
\eta_2 &=& \eta^2_1 \left[ \frac{(\mu-2)\Psi}{2(\mu-1)} + \frac{1}{2\mu}
+ \frac{2}{(\mu-1)} - \frac{3\mu}{4(\mu-1)^2} \right] \\
\chi_{\sigma \, 1} &=& \chi_{\pi \, 1} ~=~ - \, \frac{\mu\eta_1}{2(\mu-1)} \\
\chi_{\sigma \, 2} ~=~ \chi_{\pi \, 2} &=& - \, \frac{\mu\eta^2_1}{8(\mu-1)^2}
\left[\frac{}{} 3\mu(\mu-1)\Theta + 2(\mu-2)\Psi \right. \nonumber \\
&&+~ \left. \frac{(5\mu-1)(2\mu^2-5\mu+4)}{(\mu-1)} \right] \\
\lambda_1 &=& - \, (2\mu-1)\eta_1
\end{eqnarray}
Moreover,
\begin{eqnarray}
\Pi_{B2} &=& 8z^3 [ 2(\Pio_{2B2} + \Pi_{4B} - 2z_1( 2\Pio_{5B1}
+ \Pio_{5B2} + 2\Pio_{6B1}) \nonumber \\
&&-~ (\Pio_{2B1} + \Pio_{3B} - 8z_1\Pio_{6B2}) ] \\
\Pi_{C2} &=& 24z^3[ \Pio_{2B1} + \Pio_{3B} - 2z_1( 2\Pio_{5B1}
+ \Pio_{5B2} - 2\Pio_{6B1})] \\
\Lambda_{B2} &=& 8z^3[ \Pio_{2B1} + \Pio_{3B} - 2z_1(2\Pio_{5B1}
+ \Pio_{5B2} - 2 \Pio_{6B1})] \\
\Lambda_{C2} &=& 8z^3 [ \Pio_{2B1} + 2\Pio_{2B2} + \Pio_{3B} + 2\Pio_{4B}
\nonumber \\
&&-~ 8z_1(2\Pio_{5B1} + \Pio_{5B2} - \Pio_{6B2})]
\end{eqnarray}
from which it is easy to verify that
\begin{equation}
\Pi_{B2} + \Pi_{C2} - \Lambda_{B2} - \Lambda_{C2} ~=~ 0
\end{equation}
which implies that the (23) element of the transformed determinant is zero
at $O(1/N^2)$ as at leading order and as in the $U(1)$ $\times$ $U(1)$ case,
\cite{26}, which again reinforces our point of view on the importance of
the chiral symmetry at $O(1/N^2)$. Hence the formal correction to (3.13) is
\begin{equation}
[r(\alpha-1)s(\alpha-1)-4z]\left[ \frac{p(\beta)q(\beta)}{N} - 4z^2\Pi_{1B}
+ \Pi_{B2} + \Pi_{C2} \right] ~=~ 32z^2
\end{equation}
With the values given earlier, some tedious algebra leads to
\begin{eqnarray}
\lambda_2 &=& \frac{\mu\eta^2_1}{(\mu-1)}\left[
\frac{(6\mu^2-12\mu+5)}{2(\mu-2)^2\eta_1}
- \frac{\mu(2\mu-3)(\Phi+\Psi^2)}{2(\mu-2)} \right. \nonumber \\
&+& \left. \frac{3\mu(2\mu-3)(2\mu-1)\Theta(\mu)}{8(\mu-2)} + \frac{5}{2\mu}
- \frac{1}{2\mu^2} - \frac{5}{8(\mu-2)^2} \right. \nonumber \\
&+& \left. \frac{1}{4(\mu-1)^2} - \frac{37\mu}{4} + \frac{23}{16(\mu-1)}
+ 4\mu^2 + \frac{3}{8} - \frac{55}{16(\mu-2)} \right. \nonumber \\
&+& \left. \Psi(\mu) \left( 3 - \frac{1}{\mu} + \frac{3}{8(\mu-1)}
+ \frac{25\mu}{4} - 4\mu^2 \right. \right. \nonumber \\
&&+~ \left. \left. \frac{\mu(2\mu-3)}{8} \left( \frac{5}{(\mu-2)^2}
- \frac{4}{(\mu-2)} + \frac{9}{(\mu-1)} \right) \right) \right]
\end{eqnarray}
Also, from (5.13) one can restrict to three dimensions to find
\begin{equation}
\lambda ~=~ \frac{1}{2} - \frac{16}{3\pi^2N}
+ \frac{8(27\pi^2+125)}{27\pi^4N^2}
\end{equation}
Indeed, it is worth comparing (5.14) to the $O(1/N^2)$ versions of $\lambda$
in other four fermi models to observe a reassuring similarity in their
structure. For instance, in the $O(N)$ model, \cite{14,15},
\begin{equation}
\lambda ~=~ \frac{1}{2} - \frac{16}{3\pi^2N} + \frac{32(27\pi^2+632)}
{27\pi^4N^2}
\end{equation}
and in the $U(1)$ $\times$ $U(1)$ case, \cite{26},
\begin{equation}
\lambda ~=~ \frac{1}{2} - \frac{16}{3\pi^2N}
+ \frac{16(27\pi^2+472)}{27\pi^4N^2}
\end{equation}
Thus, numerically the exponents differ first at $O(1/N^2)$ in each model.

To conclude with we remark that we have discovered that the first occurence of
distinct behaviour between the $M$ $=$ $2$ and $M$ $>$ $2$ models arises at
$O(1/N^2)$ which could therefore not be noticed in a canonical large $N$
approach. This feature is intimately related to the equivalence or otherwise
of the anomalous dimensions of the $\sigma$ and $\pi$ fields, and therefore the
realisation of a chiral symmetry. However, the deeper implications that this
important observation has in relation to four dimensional field theories is
beyond the scope of the present paper but does deserve consideration.

\vspace{1cm}
\noindent
{\bf Acknowledgement.} The author thanks Simon Hands for useful communications.

\newpage
\noindent
{\Large {\bf Figure Captions.}}
\begin{description}
\item[Fig. 1.] Leading order skeleton Dyson equations with dressed
propagators.
\item[Fig. 2.] Additional graphs for $\lambda_1$.
\item[Fig. 3.] Higher order corrections for $\psi$ equation.
\item[Fig. 4.] $O(1/N^2)$ contributions to $\sigma$ equation.
\end{description}
\end{document}